\documentclass[aps,prb,reprint,twocolumn,superscriptaddress]{revtex4-1}
\usepackage{amsmath}
\usepackage{amsfonts}
\usepackage{graphicx}
\usepackage{dcolumn}
\usepackage{bm}
\usepackage{amsmath}
\usepackage{hyperref}
\usepackage{float}
\usepackage{natbib}
\usepackage{multirow}
\providecommand{\nex}{ {n_{\rm {ex}}} }

\begin{document}

\title{Unidimensional model of the ad-atom diffusion on a substrate submitted to a standing acoustic wave I. Derivation of the ad-atom motion equation}
\author{N. Combe}
\affiliation{CNRS ; CEMES (Centre d'Elaboration des Mat\'eriaux et d'Etudes Structurales); BP 94347, 29  rue J. Marvig, F-31055 Toulouse, France.}
\affiliation{Universit\'e de Toulouse ; UPS ; F-31055 Toulouse, France}

\author{C. Taillan}
\affiliation{CNRS ; CEMES (Centre d'Elaboration des Mat\'eriaux et d'Etudes Structurales); BP 94347, 29  rue J. Marvig, F-31055 Toulouse, France.}
\affiliation{Universit\'e de Toulouse ; UPS ; F-31055 Toulouse, France}

\author{J. Morillo}
\affiliation{CNRS ; CEMES (Centre d'Elaboration des Mat\'eriaux et d'Etudes Structurales); BP 94347, 29  rue J. Marvig, F-31055 Toulouse, France.}
\affiliation{Universit\'e de Toulouse ; UPS ; F-31055 Toulouse, France}
\begin{abstract}
The effect of a standing acoustic wave on the diffusion of an ad-atom on a crystalline surface is theoretically studied. 
We used an unidimensional space model to study the ad-atom+substrate system. 
The dynamic equation of the ad-atom, a Generalized Langevin equation, is analytically derived from the full Hamiltonian of the  ad-atom+substrate system submitted to the acoustic wave.
A detailed analysis of each term of this equation, as well as of their properties, is presented. 
Special attention is devoted to the expression of the effective force induced by the wave on the ad-atom. It has essentially the same spatial and time dependences as its parent standing acoustic wave. 
\end{abstract}
\date{\today}
\maketitle
\section{Introduction}

While the semi-conductors industry extensively uses the lithography process to stamp the micro-devices at the nanoscale,  research centers and laboratories have investigated the self-assembling properties of materials  to avoid this expensive and time consuming process. Most strategies to self-assemble materials at the nanoscale,  especially during the atomic deposition process of semi-conductors benefit from the elastic properties or from the structure of the substrate: the Stranski-Krastanov growth mode relies on the competition between the surface  and elastic energies to organize the 3D-growth;\cite{pimpinelli_villain,Ross1998} buried dislocations networks in the substrate induce a periodic strain field at the substrate surface that drives the diffusion of ad-atoms;\cite{Brune1998,Leroy2005} and finally the use of patterned substrates(vicinal surfaces, holes or mesas) can create some preferential nucleation sites.\cite{zhong2004,turala2009,jin1999,Mohan2010}

An alternative approach to self-assemble materials at the nano-scale, \textit{the dynamic substrate structuring effect} has been recently proposed.\cite{Taillan2011}  At the macroscopic scale, a sand bunch on a drum membrane excited at one of its eigenfrequencies self-structures by accumulating around the nodes or anti-nodes displacements of the membrane.\cite{Aranson2006} Transposing this concept at  the nanoscale, we  investigate the diffusion of an ad-atom on a crystalline substrate submitted to a standing acoustic wave (StAW).\footnote{The abbreviation SAW  is usually used to design a Surface Acoustic Wave, we thus introduce here a distinct abbreviation, StAW for Standing Acoustic Wave}
 Molecular Dynamic simulations have evidenced that  the StAW structures the diffusion of the ad-atom  by encouraging  its presence in the vicinity of the maximum
displacements of the substrate.\cite{Taillan2011}
The typical and relevant StAW wavelengths vary from few to hundreds of nanometers.  Experimentally, the production of standing surface acoustic waves of a few hundred nanometers to microns wave lengths are nowadays available through the use of interdigital transducer\cite{Sogawa2007,Takagaki2004} or optically excited nanopatterned surfaces,\cite{Siemens2009} whereas one does not know yet how to efficiently generate smaller wavelengths (few to tens nanometers) phonons. 
   
In this study, we propose to analytically  study the diffusion of a single ad-atom on a crystalline surface submitted to a StAW. 
The goal of this study is to establish the formalism and the dynamic equation that describes the diffusion of an ad-atom on a crystalline substrate submitted to a StAW.  
In section~\ref{sec:ad-atom-eq}, a generalized Langevin equation governing the ad-atom diffusion on a one-dimensional substrate is analytically derived from the Hamiltonian of the system (ad-atom+substrate).
Sects.~\ref{sec:SAW}, ~\ref{sec:memory_kernel},~\ref{sec:stochastic_force} and ~\ref{sec:potential} detail the different terms involved in this generalized Langevin equation, as well as their properties.

\section{ad-atom motion equation}
\label{sec:ad-atom-eq}
We consider the diffusion of an ad-atom on a crystalline substrate submitted to a StAW with a wave-vector in the x-direction. Since the ad-atom diffusion is expected to be mainly affected in the x direction, we specialize to a system with one degree of freedom. The extension to a 2D system to model a more complex StAW system (for instance, two StAWs with wave vectors in the x and y directions form a square lattice of nodes and anti-nodes) is straightforward, though analytical calculations may become tedious.
\begin{figure}
\begin{center}
\includegraphics[width=0.8\columnwidth]{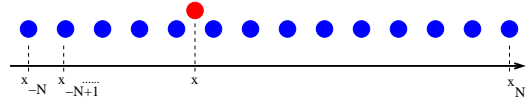}
\caption{ Schematic representation of the model under study. The ad-atom (red) and substrate atoms (blue) are characterized by their coordinates $x$ and $x_j$ ($j \in \{-N..N\}$) in the reference frame of the center of mass of the substrate. Note that, for clarity reasons, the ad-atom is not reported on the same horizontal line as the substrate atoms, but the model is unidimensional.}\label{fig1}
\end{center}
\end{figure}
Fig.~\ref{fig1} reports a sketch of the model under study. $x$  and $x_{_{-N}}, ..., x_{_{N}}$ respectively design the positions of the ad-atom and of the  $2N+1$ substrate atoms in the reference frame of the center of mass of the substrate. 
Following the work of Zwanzig~\cite{Zwanzig1973} and related works,\cite{caldeira1983,Cortes1985,Kantorovich2008} we start with the Hamiltonian of the isolated system (ad-atom+substrate): 
  \begin{equation} 
H_0 = \frac{p^2}{2m} + \Phi(x,x_{_{-N}},....,x_{_{N}}) + \sum_{j=_{-N}}^{N} \frac{p_j^2}{2m_j} + V_{\rm sub}(x_{_{-N}},...,x_{_{N}}), \label{eq:H0}
\end{equation}
 where  $m,p$ and $m_j,p_j$  are respectively the masses and momenta  of the ad-atom and of the substrate atoms, $V_{\rm sub}(x_{_{-N}},...,x_{_{N}})$ and  $\Phi(x,x_{_{-N}},....,x_{_{N}})$ the potential enegies of the substrate-substrate and ad-atom-substrate interactions.
At this point, the generation process of the StAW has not been yet introduced, this will be done later on. 

The motion of the substrate atoms will be described in the harmonic approximation\cite{kittel} with the associated phonons of eigenvibration frequencies $\omega_n$, normal coordinates $Q_n$ and momenta $\Pi_n$. 
\begin{equation} 
\sum_{j} \frac{ p_j ^2}{2m_{j}} + V_{\rm sub}( x_{_{-N}},....,x_{_{N}} ) \simeq \frac{1}{2} \sum_{n} \left[\Pi_n \bar{\Pi}_n + \omega_n^2 Q_n \bar{Q}_n\right],\label{eq:Hphonon} 
\end{equation} 
where over-bar quantities are complex conjugate quantities and where the potential origin has been fixed at the equilibrium positions, $V_{\rm sub}( x_{_{-N}}^0,....,x_{_{N}}^0)=0$. In this equation and in all the following equations, unless otherwise stated, the summations over the substrate atoms $j$ are from $-N$ to $N$,  and those over the normal modes $n$ from $-N$ to $N$ excluding $n=0$.
 Note that, within the harmonic approximation, for an isolated substrate there is no substrate dilation with temperature nor energy exchanges between the phonons.

The substrate atom displacements, $u_j = x_j - x_j^0$, around their equilibrium positions, $x_j^0$, are thus given by:
\begin{equation}
	u_j =  \frac{1}{\sqrt{m_j}} \sum_n  e^{ i k_n x_j^0 } Q_n, \label{eq:uphonon}  
\end{equation}
where $k_n$ is the wave vector of the $n$ normal mode.
In Eqs.~\eqref{eq:Hphonon} and ~\eqref{eq:uphonon} we have:
\begin{equation}
 k_{-n} =-k_n, \, \omega_{-n}=\omega_n, \, \bar{Q}_n = Q_{-n}, \,\text{and} \,\bar{\Pi}_n = \Pi_{-n}. \label{eq:symQn}
\end{equation}
From Eqs.~\eqref{eq:Hphonon} and ~\eqref{eq:uphonon} and performing a developement of the potential $\Phi$ to first order in the $u_j$'s: 
\begin{eqnarray}
\Phi(x,x_{_{-N}},....,x_{_{N}}) &=&  \Phi(x,x_{_{-N}}^0,....,x_{_{N}}^0) \nonumber \\
	&&+ \sum_j   u_j. \frac{\partial  \Phi}{\partial x_j}(x,x_{_{-N}}^0,....,x_{_{N}}^0)  \nonumber\\
	&=& \Phi_0(x) \nonumber   \\
	&&+ \frac{1}{2}\sum_n \left[ Q_n. \Psi_n(x)  + \bar{Q}_n. \bar{\Psi}_n(x) \right] \label{eq:phi}
\end{eqnarray} 
where 
\begin{eqnarray}
\Phi_0(x) &=& \Phi(x,x_{_{-N}}^0,....,x_{_{N}}^0), \label{eq:phi0}\\
\Psi_n(x) &=&  \sum_j \frac{1}{\sqrt{m_j}} e^{ i k_n x_j^0 } \frac{\partial  \Phi}{\partial x_j}(x,x_{_{-N}}^0,....,x_{_{N}}^0). \label{eq:Psi}
\end{eqnarray}The interaction of the ad-atom with the substrate has  been separated in two contributions.
 $\Phi_0(x)$, the first one, appears as an external static force field. It is due to the  frozen equilibrated substrate inter-atomic periodic potential.
The second one,  represents the interaction of the ad-atom with the phonons $Q_n$, i. e. with the moving substrate atoms around their equilibrium positions.\\
Eq.~\eqref{eq:H0} hence writes 
\begin{eqnarray}
H_0&=& \frac{p^2}{2m}  + \Phi_0(x)  \nonumber \\
&& + \frac{1}{2}\sum_n \left[ Q_n. \Psi_n(x)  + \bar{Q}_n. \bar{\Psi}_n(x) \right]\nonumber \\
&&+ \frac{1}{2} \sum_{n} \left[\Pi_n \bar{\Pi}_n + \omega_n^2 Q_n \bar{Q}_n\right],  \label{eq:H0phonon}
\end{eqnarray}
Note that the coupling between the substrate and the ad-atom is linear in the phonon variables and non-linear in the ad-atom variable, i.e. the reverse situation of the one studied by Cortes {\it et al}.~\cite{Cortes1985}

To model the presence of  a StAW in Eq.~\eqref{eq:H0phonon},   we add a forcing term with the same F  amplitude on two specific normal variables of opposite wave vectors $Q_\nex$  and $\bar Q_\nex (= Q_{-\nex})$.
However, since our model does not consider any dissipation of the substrate vibration modes, we slightly detune the forcing frequency $\Omega_\nex=\omega_\nex+\delta \omega_\nex$  from the eigenfrequency $\omega_\nex$ to avoid any resonance and subsequent divergence  of the amplitude of the mode $Q_\nex$.
These two modes will be equally excited and thus, from basic forced oscillation theory,\cite{Berkeley} one expects a forced oscillation substrate displacement field proportional to that of the parent standing wave:
\begin{equation}
 u(x,t) = - \frac{2F}{M\Delta^2} \cos[\Omega_\nex t]\cos (k_\nex x+\eta), \label{eq:standingwave}
\end{equation}
where $M$ is the mass of the oscillator, $\eta$ a phase depending on the initial conditions and with: 
\begin{equation}
\Delta^2 = \Omega_\nex^2 - \omega_\nex^2=(\omega_\nex + \delta \omega_\nex)^2 - \omega_\nex^2. \label{eq:Delta} \\
\end{equation}
 We thus consider the following Hamiltonian for the system (ad-atom+substrate submitted to a StAW):
\begin{eqnarray}
H&=& \frac{p^2}{2m}  + \Phi_0(x)  + \frac{1}{2}\sum_n \left[ Q_n. \Psi_n(x) + \bar{Q}_n. \bar{\Psi}_n(x) \right] \nonumber \\
&&+ \frac{1}{2} \sum_{n} \left[\Pi_n \bar{\Pi}_n + \omega_n^2 Q_n \bar{Q}_n\right]\nonumber \\
&& -  ( Q_\nex + \bar Q_ \nex ) F \cos[\Omega_\nex t ]  \label{eq:Hcomplet} 
\end{eqnarray}
Note that, in Eq.~\eqref{eq:Hcomplet}, the addition of the StAW term makes the Hamiltonian time-dependent. In addition, 
 the work of the operator to induce the StAW (the last term of Eq.~\eqref{eq:Hcomplet}) is not null on average and  leads to a monotonous increase of the average energy of the system (ad-atom + substrate).
 This would be the case even taking into account all the nonlinear terms we have omitted in Eq.~\eqref{eq:Hcomplet}. 
 We however assume that despite this monotonous increase of the energy, the temperature of the system remains constant, either by considering that the substrate is infinite and has the behavior of a thermostat, or by considering that the system is not totally isolated and coupled to an external thermostat. \\

The dynamic equations derived from Eq.~\eqref{eq:Hcomplet} read:\cite{Maradudin1971} 
\begin{subequations}
\begin{align} 
\frac{d Q_n}{dt}   &= \Pi_n   && \label{eq:hamilt_sub0}\\
\frac{d \Pi_n}{dt} & =- \omega_n^2 Q_n - \bar{\Psi}_n(x)    + \Lambda_{n,\nex} F \cos[ \Omega_\nex t ]& &\label{eq:hamilt_sub1} \\
                             \nonumber \\
\frac{d x}{dt} &  = \frac{p}{m}   && \label{eq:hamilt_ada0}\\
\frac{d p}{dt} &= - \frac{d \Phi_0}{d x}(x)  - \frac{1}{2}\sum_n \left[ Q_n \frac{d\Psi_n }{d x}(x)  + \bar{Q}_n \frac{d \bar{\Psi}_n }{d x}(x)  \right] & &\label{eq:hamilt_ada1}
\end{align}
\end{subequations}
Where $\Lambda_{i,j} = \delta_{i,j} + \delta_{i,-j}$ with  $\delta_{i,j}$ the Kronecker symbol\footnote{ $\delta_{i,i}=1$ and $\delta_{i,j}=0$ if $i \neq j$}. 
 In Eq.~\eqref{eq:hamilt_sub1}, $- \bar{\Psi}_n(x)$ is the force  on the substrate normal mode  $n$, induced by the ad-atom at position $x$. 
Solving Eq.~\eqref{eq:hamilt_sub0} and Eq.~\eqref{eq:hamilt_sub1} between $t_0$ and $t$, the normal substrate coordinates read:   
\begin{widetext}
\begin{equation} 
Q_n(t) = Q_n(t_0) \cos[\omega_n (t-t_0)] +  \frac{ \Pi_n(t_0) }{\omega_n} \sin[ \omega_n (t-t_0) ]
- \int_{t_0}^{t}   \bar{\Psi}_n(x(t')) \frac{\sin[ \omega_n(t-t') ]}  { \omega_n }dt'  
  + \int_{t_0}^{t}   \Lambda_{n,\nex} F \cos[\Omega_\nex t' ] \frac{\sin[ \omega_n(t-t') ]}{ \omega_n} dt',\label{eq:Qnt1}
\end{equation} 
\end{widetext}
where  $Q_n(t_0)$ and $\Pi_n(t_0)$ are fixed by the initial conditions.
An integration of the second integral and an integration by parts of the first one gives:
\begin{eqnarray} 
Q_n(t)&= &C_n(t_0)  \cos[\omega_n (t-t_0)]  +  D_n(t_0) \sin[\omega_n (t-t_0)]\nonumber \\ 
& - & \frac{\bar{\Psi}_n(x(t)) }{\omega_n^2} \nonumber  \\ 
&+& \int_{t_0}^t \frac{\cos [\omega_n (t-t')]}{\omega_n^2} \frac{dx}{dt}(t') \frac{d\bar{\Psi}_n}{d x}(x(t'))  dt' \nonumber \\ 
&-& \Lambda_{n,\nex} \frac{F}{\Delta^2 } \cos[\Omega_\nex t],  \label{eq:Qnt2}
\end{eqnarray} 
with:
\begin{subequations}
\begin{align}
C_n(t_0) &=& Q_n(t_0) + \Lambda_{n,\nex} \frac{ F}{\Delta^2}\cos[\Omega_\nex t_0] + \frac{\bar{\Psi}_n(x(t_0))}{\omega_n^2}    \label{eq:Cnt0} \\
D_n(t_0) &=& \frac{ \Pi_n(t_0) }{\omega_n} - \Lambda_{n,\nex} \frac{ F}{\Delta^2} \frac{ \Omega_\nex}{\omega_n} \sin[\Omega_\nex t_0] \label{eq:Dnt0}.
\end{align}
\end{subequations}
From Eqs.~\eqref{eq:symQn} and \eqref{eq:Psi} we have:
\begin{equation}
 \bar C_n= C_{-n} \quad \text{and}  \quad \bar D_n= D_{-n} \label{eq:symCn}
\end{equation}
Using  Eq.~\eqref{eq:hamilt_ada0},~\eqref{eq:hamilt_ada1} and ~\eqref{eq:Qnt2}, we derive the generalized Langevin equation governing the ad-atom diffusion:
\begin{eqnarray}
m \frac{d^2 x} {dt^2}\nonumber &=&  - \frac{d \Phi_{\text{eff}}}{d x}(x)    
-   \int_{t_0}^t \gamma(x(t),x(t'),t-t') \frac{dx}{dt}(t')  dt' + \xi(t)  \nonumber \\
&&	  +   F_{SAW}(x,t).   \label{eq:langevin_almost_end} 
\end{eqnarray}
The left hand side term of Eq.~\eqref{eq:langevin_almost_end} is the usual inertial term.
On the right hand side, we distinguish four terms, which are successively:
\begin{itemize}
	\item  the force induced by the effective crystalline potential $\Phi_{\text{eff}}(x)$, defined by: 
	\begin{equation} 
	\Phi_{\text{eff}}(x) =  \Phi_0(x)   -  \frac{1 }{ 2} \sum_n \frac{1 }{ \omega_n^2}  \Psi_n(x) \bar{\Psi}_n(x). \label{eq:phieff}
	\end{equation} 
	The properties of this potential will be studied in Sect.~\ref{sec:potential}.
	\item  The friction term  $-\int_{t_0}^t \gamma(x(t),x(t'),t-t') \frac{dx}{dt}(t')  dt'$ that depends on the ad-atom velocity and on the memory kernel $\gamma(x,x',t-t')$:
	\begin{equation}
	 \gamma(x,x',t-t') = \sum_n    \frac{ \cos ( \omega_n (t-t')  )  } { \omega_n^2}   \frac{d \Psi_n}{d x}( x )  \frac{d \bar{\Psi}_n}{d x}( x' ). \label{eq:gam_init}
	\end{equation}
	The properties of $\gamma(x,x',t-t')$ will be studied in Sect.~\ref{sec:memory_kernel}.
	\item  the stochastic force~\cite{Kantorovich2008,Cortes1985} $\xi(t)$:
	\begin{eqnarray}
		\xi(t) &=& -  \sum _n \bigg[ C_n(t_0) \cos[ \omega_n (t-t_0)] \nonumber \\
		&& +   D_n(t_0) \sin[ \omega_n (t-t_0)]  \bigg]  \frac{d \Psi_n}{d x} (x(t)).  \label{eq:Csi}
	\end{eqnarray} 
	This term depends on the initial conditions and ad-atom position and is a quickly varying force generated by the substrate. The properties of this force will be described in Sect.~\ref{sec:stochastic_force}.
	\item The last term $F_{SAW}(x,t)$ is the effective force due to the applied forcing term at $\Omega_\nex$, i.e. the  force $F_{SAW}(x,t)$ induced by the StAW on the ad-atom through the substrate: 
	\begin{equation}
	F_{SAW}(x,t) = \frac{F}{\Delta^2 } \bigg(\frac{d \Psi_\nex}{d x} (x) 
	+\frac{d \bar \Psi_\nex}{d x}(x) \bigg) \cos[\Omega_\nex t]. \label{eq:FSAW1}
	\end{equation}
	This force will be detailed in Sect.~\ref{sec:SAW}.
 \end{itemize} 

The three first forces, crystalline, friction and stochastic, exist even in the absence of the StAW excitation.
They are the usual forces describing the dynamics of the atoms in a crystalline material.\\
We have chosen to keep in $F_{SAW}(x,t)$ only the forced oscillation term at the pulsation $\Omega_\nex$. All the other terms depending on $F$ have been included in the stochastic force $\xi(t)$. 
They correspond to the answers of the oscillators $Q_\nex$ and  $\bar{Q}_{\nex}$ to the initial conditions at $t=t_0$. Since the normal modes of the substrate are undamped, these last terms are periodic and do not cancel. For damped oscillators, the terms depending on $F$ in the stochastic force would correspond to a transient regime and would thus cancel, contrary to the forced oscillation term at the pulsation $\Omega_\nex$.

\section{The StAw Force} 
\label{sec:SAW}

To derive the expression of the force $F_{SAW}(x,t)$ induced by the StAW, we need to explicit the expression of $\Psi_\nex(x)$ in Eq.~\eqref{eq:FSAW1}.
Since interaction potentials depend only on the relative position of the interacting particles, so do $\Phi_0$ and  $\Psi_n$.  $\Psi_n$ writes then:
\begin{eqnarray}
\Psi_n(x) &=&  \sum_j \frac{1}{\sqrt{m_j}} e^{ i k_n x_j^0 }   \frac{ \partial \Phi}{\partial x_j}(x-x_{_{-N}}^0,....,x-x_{_{N}}^0) \nonumber \\
&  = &     \alpha_n (x) e^{ i k_n x },   \label{eq:Psinex0}
\end{eqnarray}
with  $\alpha_n(x)$ defined as:
\begin{equation}
\alpha_n (x)=  \sum_j \frac{1}{\sqrt{m_j}}  e^{ -i k_n  (x-x_j^0) }   \frac{ \partial \Phi}{\partial x_j} (x-x_{_{-N}}^0, ..., x-x_{_{N}}^0) \label{eq:alpha0}
\end{equation}
Note that for an infinite crystal the $\alpha_n(x)$ functions have the lattice periodicity.\footnote{The periodicity of the $\alpha_n$ functions can also be related to the Bloch function character of the $\Psi_n$ functions}. In addition,  $\bar{\alpha}_n(x) = \alpha_{-n}(x)$ so that introducing the real $\alpha^r_n(x) = \Re(\alpha_n(x))$ and imaginary $\alpha^i_n(x) =\Im(\alpha_n(x))$ parts of $\alpha_n(x)$, we have: 
\begin{equation} 
\alpha^r_n(x) = \alpha^r_{-n}(x) \quad  \text{and} \quad \alpha^i_n(x) = - \alpha^i_{-n}(x), \label{eq:sym_alpha}
\end{equation} 
which leads to:
\begin{equation}
\frac{d \Psi_n}{d x} + \frac{d \bar \Psi_n}{d x}= 2[g_n(x)  \cos(k_n x) + h_n(x) \sin( k_n x)], \label{eq:dPsi} \\
\end{equation}
with
\begin{equation}
g_n =  \frac{d \alpha^r_n}{dx}- k_n \alpha^i_n \quad \quad
h_n =  -  (k_n \alpha^r_n + \frac{d \alpha^i_n}{dx}). \label{eq:geth}
\end{equation}
where $g_n(x)$ and $h_n(x)$ have the lattice substrate periodicity. 
The $F_{SAW}(x,t)$ force writes then:
\begin{widetext}
\begin{equation}
F_{SAW}(x,t)= \frac{2F}{\Delta^2 }  \cos(\Omega_\nex t) \left[  g_\nex(x)  \cos(k_\nex x) + h_\nex(x) \sin( k_\nex x)     \right]  \label{eq:FSAW2}\\ 
\end{equation}
\end{widetext}
 
The comparison of Eq.~\eqref{eq:FSAW2} to Eq.~\eqref{eq:standingwave} shows that, as expected, the SAW force on the ad-atom, induced by the standing surface acoustic wave through the substrate, has the large scale spatial and time dependence of the corresponding standing wave. 
This dependence at $2\pi/k_\nex$ scale has been exhibited in  Molecular Dynamic simulations~\cite{Taillan2011} of ad-atom diffusing on a substrate submitted to a standing surface acoustic wave.
However, at a finer scale, $x$ smaller than the lattice parameter, this force experiences an amplitude and a phase modulation due to the presence of the crystalline potential through the functions $g_n(x)$ and $h_n(x)$.\\
At this point, it is instructive to turn to a particular case by specifying the substrate and  the interaction potential between the ad-atom and the substrate atoms, especially in order  to get an explicit expression of the functions $\alpha_n(x)$ and thus of  $g_n(x)$ and $h_n(x)$. We assume that  the substrate atoms have the same mass $M$ and that
 the ad-atom interacts with each substrate atom through an attracting pair potential $V_{\rm{pair}}(x-x_i)$ that cancels at infinity.  We choose for $V_{\rm{pair}}$ an exponential curve of extension $\sigma$ (roughly the pair interaction range), i.e. a potential expression, that is physically meaningful and that allows the derivation of  analytical calculations. 
\begin{equation} 
\Phi(x,x_{_{-N}}^0...x_{_{N}}^0) = \sum_j V_{\rm {pair}}(x-x_j^0)=  - \sum_j V_0 e^{ \frac{- |x-x_j^0|} {\sigma}} \label{eq:pot_phi}
\end{equation} 
Where $V_0$ is the bonding energy. Note that  minima of  $\Phi$ correspond to atoms substrate positions.  We have $x_j^0 = j a$ where  $a$ is the lattice spacing and $j \in [-N,N]$. $\alpha_n (x)$ then writes:  
\begin{eqnarray}
\alpha_n (x) & =& \frac{1}{\sqrt{M }}  \sum_j e^{ -i k_n (x-x_j^0) }   \frac{ \partial \Phi}{\partial x_j} (x-x_{_{-N}}^0, ..., x-x_{_{N}}^0)  \nonumber \\
&=&- \frac{1}{\sqrt{M }}  \sum_j  e^{ -i k_n (x-x_j^0) }   \frac{d V_{\rm{pair}}}{d x} (x-x_j^0)\nonumber \\
&=& \frac{V_0}{\sqrt{M }} \sum_{j}  e^{ -i k_n (x-j a) }   \frac{ d }{d x}  \left[ e^{ \frac{- |x-ja|} {\sigma}} \right] \label{eq:alpha1}
\end{eqnarray}
To take into account the discontinuties of the derivative of $V_{\rm {pair}}$ at its minima, we define $m_0(x)$ and $r(x)$ respectively, the quotient and  the rest of the Euclidian division of $x$ by $a$: $x = m_0 a + r$, with $m_0 \in [-N,+N]$ and $0 \le r(x) <a$. 
$m_0(x)$ tells us in between which potential wells $[m_0a, (m_0+1)a]$  the ad-atom is and $r(x)$ where it is exactly in between.
Extending the size of the substrate to infinity ($N \to \infty$) in Eq.~\eqref{eq:alpha1}, we obtain:
\begin{eqnarray}
\lefteqn{ \alpha_n(x)  
 =\frac{V_0}{\sigma \sqrt{M }}  \left[ \sum_{j=m_0+1}^{\infty} e^{ -i k_n  (x-ja) }  e^{ \frac{x-j a}{\sigma} }  \right. } \nonumber\\
&& -  \left. \sum_{j=-\infty}^{m_0}  e^{ -i k_n  (x-ja) }e^{-  \frac{(x-ja)} {\sigma}} \right] \label{eq:alpha2}\\
&=&\frac{V_0}{\sigma \sqrt{M }}  \left[   \frac{ e^{ - i k_n  r + \frac{ r} {\sigma} }} { e^{  \frac{a} {\sigma} - i k_n a } -1 } \right.  - \left.  \frac{e^{  -i k_n r - \frac{ r} {\sigma} }} { 1 - e^{ - i k_n a-\frac{a} {\sigma}} } \right]\label{eq:alpha3}
\end{eqnarray}
$\alpha_n(x)$ appears then as a function of $r(x)$ only, which writes:
\begin{equation}
\alpha_n (r(x)) = \frac{V_0}{\sigma \sqrt{M }} \frac{ e^{ik_n a} \cosh( \frac{r}{\sigma} )  - \cosh(\frac{r-a}{\sigma} ) } { \cosh( \frac{a}{\sigma}) - \cos(k_n a) } e^{ - i k_n r }\label{eq:alpha4}
\end{equation}

From this expression of $\alpha_n$ we deduce the following expressions for $\Psi_n$, $g_n$ and $h_n$:
\begin{widetext}
\begin{eqnarray}
\Psi_n(x) &=&  \frac{V_0}{\sigma \sqrt{M }} \frac{ e^{ik_n a} \cosh( \frac{r}{\sigma} )  - \cosh(\frac{r-a}{\sigma} ) } { \cosh( \frac{a}{\sigma}) - \cos(k_n a) } e^{  i k_n  m_0 a }, \label{eq:Psinex}  \\
g_n(r) &=&  \frac{V_0 }{\sigma^2 \sqrt{M} [ \cosh( \frac{a} {\sigma})-\cos(k_n a) ]}  \left[ \cos[(k_n(r-a)] \sinh ( \frac{r} {\sigma} ) -  \cos (k_n r)\sinh(\frac{r-a}{\sigma}) \right], \label{eq:g}\\
h_n(r) &=&  \frac{V_0 }{\sigma^2 \sqrt{M} [ \cosh( \frac{a} {\sigma})-\cos(k_n a) ]}  \left[ \sin[(k_n(r-a)]\sinh ( \frac{r} {\sigma} ) - \sin(k_n r) \sinh (\frac{r-a}{\sigma}) \right].\label{eq:h}
\end{eqnarray}
\end{widetext}
Note that, since $g_n(x)$ and $h_n(x)$ in Eq.~\eqref{eq:FSAW2} have the lattice substrate periodicity, we have $g_n(x) =g_n(m_0 a + r)= g_n(r)$ and $h_n(x) =h_n(m_0 a + r)= h_n(r)$. One can easily verify that $F_{SAW}$ (Eq.~\eqref{eq:FSAW2}) is a continuous function of $x$, despite the discontinuity of the derivative of $V_{\rm {pair}}$. 
A more symetric expression can be obtained through the $r=r' + a/2$ translation,  with now $-a/2 \leq r' \leq a/2$ ($r'=0$ corresponds to the mid position between two successive potential wells, located at $r'=\pm a/2$):
\begin{widetext}
\begin{equation} 
  F_{SAW}(x,r'(x),t)= F_{\rm {saw}}(r')  \cos (\Omega_\nex t) \cos[k_\nex(x-r') + \varphi_0(r')]= F_{\rm {saw}}(r')  \cos (\Omega_\nex t) \cos[k_\nex x + \varphi(r')]\label{eq:Fsaw_norm2},\\
\end{equation}
with 
\begin{eqnarray} 
F_{\rm {saw}}(r') &=&  2 F_0 \left[ \cos^2\frac{k_\nex a}{2}\sinh^2\frac{ a}{2\sigma}\cosh^2\frac{r'}{\sigma} + \sin^2\frac{k_\nex a}{2}\cosh^2\frac{ a}{2\sigma}\sinh^2\frac{ r'}{\sigma}\right]^{1/2}, \nonumber \\
 &=& 2 F_0 \cos(\frac{k_\nex a}{2}) \sinh(\frac{a}{2\sigma}) \left[ 1+ \left( 1 + \tan^2\frac{k_\nex a}{2}\coth^2\frac{ a}{2\sigma} \right) \sinh^2\frac{ r'}{\sigma}\right]^{1/2}, \label{eq:Fsaw1}\\
 \tan(\varphi_0(r')) &=& \tan\frac{k_\nex a}{2} \coth \frac{ a}{2\sigma}\tanh\frac{ r'}{\sigma}, \label{eq:phase}\\
  F_0 &=&  \frac{ 2 V_0 F }{\Delta^2 \sigma^2 \sqrt{M} [ \cosh( \frac{a} {\sigma})-\cos(k_\nex a) ]} 
\end{eqnarray} 
\end{widetext}
where $F_{\rm {saw}}(r')$ and $\varphi(r')=\varphi_0(r')-k_\nex r'$ are respectively the amplitude and the phase of the large scale spatial dependence of $F_{SAW}(x,r'(x),t)$. 
Eq.~\eqref{eq:Fsaw_norm2} again evidences the large scale spatial  and time dependence of the SAW.
This point is also evidenced by evaluating the force at the substrate atoms positions, $r'= \pm a/2$,  and at the midway position between two successive potential wells, $r'=0$:
\begin{widetext}
\begin{eqnarray} 
F_{SAW}(x,r'=\pm a/2,t)&=& F_0  \sinh(\frac{a}{\sigma})  \cos(k_\nex x)  \cos (\Omega_\nex t) \label{eq:ra2}\\
F_{SAW}(x,r'=0,t)&=&2  F_0  \cos(\frac{k_\nex a}{2}) \sinh(\frac{a}{2\sigma}) \cos(k_\nex x)  \cos (\Omega_\nex t)  \label{eq:r0}
\end{eqnarray}
\end{widetext}

\begin{figure}
\begin{center}
\includegraphics[width=0.8\columnwidth]{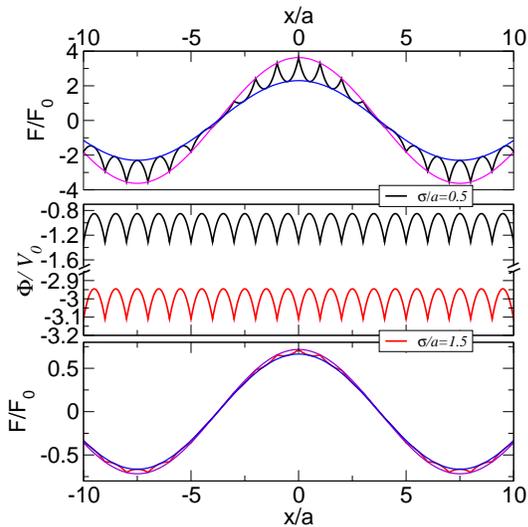}
\caption{Top and bottom: maximum force induced by the StAW ($t=0[\frac{2 \pi}{\Omega_\nex}]$ in Eq.~\eqref{eq:Fsaw_norm2}) and  midle: interatomic potential (Eq.~\eqref{eq:pot_phi}), as a function of $x/a$,  for  $k_\nex a=2 \pi/15$ and two values of $\sigma/a$: 0.5 (black) and 1.5 (red). Top and Bottom:  envelop curve at the substrate atom positions (blue, Eq.~\eqref{eq:r0}) and midway in between ( magenta, Eq.~\eqref{eq:ra2}).}\label{fig2}
\end{center}
\end{figure}
Fig.~\ref{fig2} reports both the maximum force  ($t=0[\frac{2 \pi}{\Omega_\nex}]$ in Eq.~\eqref{eq:Fsaw_norm2}) induced by the StAW  and the interatomic potential (Eq.~\eqref{eq:pot_phi})  as a function of $x/a$ for $k_\nex a=2 \pi/15$ and two values of $\sigma/a$: $0.5$  and $1.5$.  The  large scale spatial dependence in $\cos(k_\nex x)$  of the force  $F_{SAW}(x,r'(x),t)$ is  clearly evidenced, whereas the finer scale, between two successive potential wells exhibits the sinus hyperbolic-based dependence of the force evidenced in Eq.~\eqref{eq:Fsaw1}. 
As $\sigma$ increases, the amplitude of the wells of the ad-atom-substrate potential (Eq.~\eqref{eq:pot_phi}) and the amplitude of the variations of the force at both the large scale $2\pi/k_\nex$  and at the fine scale $a$ decrease:  indeed,  if the interaction between the ad-atom and the substrate is less pronounced, the force induced by the wave on the ad-atom well will also be reduced on both fine and large spatial scales.

\section{The memory kernel }
\label{sec:memory_kernel}

 Let's now study the  memory kernel $\gamma(x,x',t-t')$ of the friction force (Eq.~\eqref{eq:gam_init}) that depends on the  $\alpha_n$ functions through $ \Psi_n(x)$ Eq.~\eqref{eq:Psinex0}:
 \begin{equation}
	 \gamma(x,x',t-t') = \sum_n    \frac{ \cos ( \omega_n (t-t')  )  } { \omega_n^2}   \frac{d \Psi_n}{d x}( x )  \frac{d \bar{\Psi}_n}{d x}( x' ) \nonumber
	\end{equation}
	Note that this memory kernel depends on the ad-atom position so that the dissipation term in Eq.~\eqref{eq:langevin_almost_end} is nonlinear in the ad-atom variables.~\cite{Zwanzig1973,Lindenberg1984} 
An explicit expression of $\gamma$ is out of scope. However, since the $\alpha_n$ functions are periodic functions of period the lattice parameter $a$, we can make an evaluation of the kernel without taking into account their spatial variations. They are then replaced in Eq.~\eqref{eq:gam_init} by their mean value over the period $a$. This is equivalent to take into account only the first term, $\tilde{\alpha}_n(0)$, of their Fourrier expansion:
\begin{equation}
\gamma(x,x',t-t')\approx 
 \sum_n  \frac{ \cos[\omega_n (t-t')]e^{ i k_n (x - x')}}{\omega_n^2} 
k_n^2\tilde{\alpha}_n(0) \overline{\tilde{\alpha}_n(0)}\label{eq:gamma2},
\end{equation}
with
\begin{equation}
 \tilde{\alpha}_n(0)  =  \frac{1}{a} \int_0^a  \alpha_n(x) dx. \label{eq:alphamoyen} \\
\end{equation}
Again using the particular inter-atomic potential (Eq.~\eqref{eq:pot_phi}) with  Eq.~\eqref{eq:alpha3} or \eqref{eq:alpha4} one gets:
\begin{equation}
 \tilde{\alpha}_n(0)  = \frac{2ik_nV_0}{a\sigma \sqrt M (k_n^2 + 1/\sigma^2)} \label{eq:alphamoyen2}
\end{equation}
Within this approximation, the memory kernel reads: 
\begin{eqnarray}
 \gamma(x-x',t-t')  \approx   \frac{4 V_0^2}{a^2 M}\nonumber\\
 \sum_n     \frac{ k_n^4\cos[\omega_n (t-t')]} { \omega_n^2}&& \left[\frac{ \sigma}{ 1+  k_n^2 \sigma^2}\right]^2 e^{i k_n (x-x')}  \label{eq:gam}
\end{eqnarray}
In the same spirit, we will use the Debye model,~\cite{Ashcroft} which is well adapted for simple monoatomic lattices at intermediate temperatures, to describe the phonon dispersion relation, $\omega_n = c_s k_n$, where $c_s$ is the  speed of sound of the substrate; and change the discrete summation to an integral:
\begin{eqnarray}
 \gamma(x-x',t-t') \approx \frac{4 V_0^2\sigma^2}{a^2 M c_s^2} \nonumber\\
   \int_{-k_{D}}^{k_{D}}    \frac{ k^2 \cos[c_sk (t-t')]e^{i k(x-x')} }{ [1+  k^2  \sigma^2]^2}  &&g(k) dk,  \label{eq:gamint}
\end{eqnarray}
where $k_D=\frac{\pi}{a}$ is the Debye wave number and $g(k)=L/(2\pi)$ the density of states in the reciprocal space, with $L=2Na$ the size of the substrate. 
 Moreover, since the function $k^2/[1+  k^2  \sigma^2]^2$ is a peaked function centered at $k=0$ of extension $1/\sigma$, and considering that  $\sigma$ is generally larger than $a$, the limits of integration are extended to $\infty$.  
An integration by parts leads to the calculation of  Fourier transform of  Lorentzians and to the following approximated $\gamma$ expression:
\begin{eqnarray}
\gamma(x-x',t-t') & =& \frac{ L V_0^2 }{2c_s^2 a^2 M  \sigma} \left[   H \bigg(\frac{|x-x' + c_s(t-t')|}{\sigma} \bigg) \right. \nonumber \\ 
&&+ \left. H\bigg(\frac{|x-x' - c_s(t-t')|}{\sigma} \bigg) \right]   \label{eq:gam_fin}\\
\mbox{with}& \ &\ H(x) = (1 - x) e^{-x} \nonumber
 \end{eqnarray}
The expression of the memory Kernel in Eq.~\eqref{eq:gam_fin} is an even function of $x-x'$ and $t-t'$. The dependence on $x-x'$ is a direct consequence of the elusion of the dependence of $\alpha_n(x)$ on $x$ (at the scale $a$) (see Sect.~\ref{sec:potential}). 
  We do not find for $\gamma(x,x',t-t')$ a simple exponentially decreasing function of $|t-t'|$   as usually assumed in most textbooks~\cite{ pottier2007}.
   However, we emphasize that the $\gamma$ expression in  Eq.~\eqref{eq:gam_fin} crucially depends on the interaction potential chosen (Eq.~\eqref{eq:pot_phi}) and that Eq.~\eqref{eq:gam_fin} provides a rather crude estimation of $\gamma(x,x',t-t')$: we have ignored the dependence of $\Psi_n$ on the length scale $a$ and the extension of the integral Eq.~\eqref{eq:gamint} to infinity is a rough assumption ($\sigma/a$  is not in general very large compared to $1$). \\
   In addition, from Eq.~\eqref{eq:gam_fin} the correlation time appears to be of the order of $\sigma/c_s$. Knowing that $\sigma$ is of the order of magnitude of the lattice paramater, this correlation time is of the order of the inverse of the Debye frequency.

\section{The stochastic force}
\label{sec:stochastic_force}

In this section, we describe the properties of $\xi(t)$, the stochastic force (Eq.\eqref{eq:Csi}). Since this force  depends on the ad-atom position through the coupling term $\frac{d \Psi_n}{d x} (x(t))$, it represents multiplicative fluctuations.~\cite{Lindenberg1984}
Using Eqs.~\eqref{eq:Cnt0} and \eqref{eq:Dnt0}, it  writes: 
\begin{widetext}
\begin{eqnarray}
\xi(t) &=& -  \sum _n \bigg[ \left( Q_n(t_0) + \Lambda_{n,\nex} \frac{ F}{\Delta^2}\cos[\Omega_\nex t_0] + \frac{\bar{\Psi}_n(x(t_0))}{\omega_n^2} \right) \cos[ \omega_n (t-t_0)] \nonumber \\
		&& +  \left( \frac{ \Pi_n(t_0) }{\omega_n} - \Lambda_{n,\nex} \frac{ F}{\Delta^2} \frac{ \Omega_\nex}{\omega_n} \sin[\Omega_\nex t_0] \right) \sin[ \omega_n (t-t_0)]  \bigg]  \frac{d \Psi_n}{d x} (x(t)).  \label{eq:xito}
\end{eqnarray} 
\end{widetext}
This force partially results from the initial state of the substrate. In that sense, our system is completely deterministic.  However,  we have considered a quadratic approximation in  Eq.~\eqref{eq:Hphonon} and a linear development of $\Phi$ in Eq.~\eqref{eq:phi}. In a real substrate, the non-linear terms can hold and/or exchange some energy with the normal substrate modes and in addition the substrate is never completely uncoupled to the experimental set-up. To take into account these exchanges of energy without explicitly describing them, we characterize the state of the substrate ($\vec{Q}$ ,$\vec{\Pi}$) at $t_0$ using a probability distribution $p(\vec{Q}(t_0),\vec{\Pi}(t_0))$, where $\vec{Q}$ and $\vec{\Pi}$ are vectors whose coordinates are the variables $Q_n$ and $\Pi_n$.  We suppose that the StAW forcing terms in Eq.~\eqref{eq:Hcomplet} initially switched off are switched on at $t_0$: the Hamiltonian describing our system at $t<t_0$ is thus given by  Eq.~\eqref{eq:H0phonon}.  \\
Besides, if we want Eq.~\eqref{eq:langevin_almost_end} to be regarded as a conventional Generalized Langevin equation,  the quantity $\xi(t)$ ought to have the properties that are expected for Langevin noise. Especially,  its average is expected to cancel with respect to the probability distribution $p(\vec{Q}(t_0),\vec{\Pi}(t_0))$. \cite{zwanzig_book}
In order to satisfy this last requirement, we choose the following expression for $p(\vec{Q}(t_0),\vec{\Pi}(t_0))$: 
\begin{equation}
 p(\vec{Q}(t_0),\vec{\Pi}(t_0)) = Z^{-1} e^{-\beta H_s}, \label{eq:Boltzmann}
 \end{equation}
where $\beta=1/(k_B T)$, $k_B$ the Boltzmann constant, T, the temperature of a surrounding thermostat that mimics the coupling of the system with the experimental set-up and $H_s$ given by: 
 \begin{eqnarray} 
  H_s(\vec{Q},\vec{\Pi}) &= &
 \frac{1}{2}  \sum_n \left[\Pi_n \bar{\Pi}_n + \omega_n^2 Q_n \bar{Q}_n\right] \nonumber \\
  &&+  \frac{1}{2}\sum_n  \left[Q_n. \Psi_n(x(t_0)) +  \bar{Q}_n.\bar{\Psi}_n(x(t_0))\right] \nonumber \\
  && +  \frac{1}{2} \left[ Q_\nex + \bar{Q}_{\nex} \right] \frac{ F \omega_\nex^2}{\Delta^2}\cos[\Omega_\nex t_0]. \label{eq:H_s}
\end{eqnarray}
  $H_s$ describes the coupling between the substrate and the ad-atom at position $x(t_0)$ and contains a term derived from the StAW force to take into account the initial conditions imposed by the StAW on the $Q_n$ variables at $t=t_0$. The Hamiltonian $H_s$ is hence different  from the  $H_0$ one (Eq.~\eqref{eq:H0phonon}) of the system for $t<t_0$ i.e.  the probability distribution $p(\vec{Q}(t_0),\vec{\Pi}(t_0))$ corresponds to a non-equilibrium (macro-)state of  the system described by $H_0$ coupled to a thermostat at temperature T. We will now establish the properties of the fluctuating force $\xi(t)$ for the probability distribution Eq.~\eqref{eq:Boltzmann}.

The examination of Eqs.~\eqref{eq:xito} and \eqref{eq:H_s} reveals that the appropriate variables are:
\begin{equation}
 R_n= Q_n + \frac{\bar \Psi_n}{\omega_n^2}  + \Lambda_{n,\nex} \frac{ F}{\Delta^2}\cos[\Omega_\nex t_0]. \label{eq:Rn}
\end{equation}
With these variables $H_s$ and $\xi(t)$ write:
\begin{eqnarray}
   H_s(\vec{Q},\vec{\Pi}) &=& 
   \frac{1}{2}\sum_n \bigg[\Pi_n \bar{\Pi}_n + \omega_n^2  R_n \bar{R}_n \nonumber \\
   &&\  - \bigg| \frac{\bar \Psi_n}{\omega_n^2}  + \Lambda_{n,\nex} \frac{ F}{\Delta^2}\cos[\Omega_\nex t_0] \bigg|^2\bigg], \label{eq:H_s1}\\
\xi(t)& =& -  \sum _n  \frac{d \Psi_n}{d x} (x(t)) \bigg[\nonumber \\
&&  R_n(t_0)  \cos[ \omega_n (t-t_0)] \nonumber \\
&&+  \left( \frac{\Pi_n(t_0)}{\omega_n} -  \Lambda_{n,\nex} \frac{ F}{\Delta^2} \frac{\Omega_\nex}{\omega_n}\sin[\Omega_\nex t_0] \right) \nonumber \\
&& \sin[ \omega_n (t-t_0)]  \bigg].  \label{eq:xit1} 
\end{eqnarray}
From Eq.~\eqref{eq:Boltzmann} and~\eqref{eq:H_s1}, variables $\Pi_n$ and $R_n$ appear as complex variables with centered Gaussian distributions of variance $\beta^{-1}$.  Note however,  that since  $\bar{\Pi}_n=\Pi_{-n}$ and $\bar{R}_n = R_{-n}$, all these variables are not independent. One can easily re-write Eq.~\eqref{eq:H_s1} using  a set of 2N independent variables  $(R_n,\Pi_n)$ with $n> 0$: 
\begin{eqnarray} 
  H_s(\vec{Q},\vec{\Pi}) &=& 
   \sum_{n>0} \bigg[\Pi_n \bar{\Pi}_n + \omega_n^2  R_n \bar{R}_n \nonumber \\
   && - \bigg| \frac{\bar \Psi_n}{\omega_n^2}  + \Lambda_{n,\nex} \frac{ F}{\Delta^2}\cos[\Omega_\nex t_0] \bigg|^2\bigg]. \label{eq:H_s2}
\end{eqnarray}
 So that, for  any two variables $X$ and $Y$ $\in \{\omega_n R_n,\Pi_n\}$ ($n> 0$) , their mean values $\langle X \rangle$ are 0 and their covariances $ \langle[X-\langle X \rangle][\bar Y-\langle \bar Y \rangle]\rangle$ are  $(2/\beta)\delta_{XY}$.

From which we deduce the stochastic properties of $\xi(t)$
\begin{eqnarray}
\langle\xi(t)\rangle &=&  \frac{F}{\Delta^2} \left[ \frac{d \Psi_\nex}{d x} (x(t)) + \frac{d \bar \Psi_\nex}{d x} (x(t)) \right]\nonumber \\
&&\bigg[   \frac{ \Omega_\nex}{\omega_\nex} \sin[\Omega_\nex t_0] \sin[ \omega_\nex (t-t_0)] \bigg],  \label{eq:xi_moy} \\
C(t,t') &=& \langle [\xi(t)-<\xi(t)>] \ [\xi(t') -<\xi(t')>]\rangle \nonumber\\
  &=& \frac{1}{\beta} \sum_n \frac{\cos[\omega_n (t-t')]}{\omega_n^2}
  \left[\frac{d \Psi_n}{d x} (x(t))\frac{d  \bar{\Psi}_n}{d x} (x(t' ))\right]\nonumber\\
  &=& \frac{1}{\beta} \gamma(x(t),x(t'),t-t').
\end{eqnarray}
We recover in this last equation the fluctuation-dissipation theorem: this result is especially independent of the precise expression of the potentials $\Phi$ and $V_{\rm sub}$ in Eq.~\eqref{eq:H0} as soon as this later can be approximated by Eq.~\eqref{eq:H0phonon}.
The same result has been demonstrated in a general frame by Zwanzig.~\cite{Zwanzig1973}  
The non-null value of $\langle\xi(t)\rangle$ is related to the time-depend Hamiltonian~(Eq.\eqref{eq:Hcomplet}) and more precisely to the initial conditions that are imposed by abruptly switching on the StAW term at $t_0$.  The Hamiltonian $H_s$ Eq.~\eqref{eq:H_s1} actually takes into account the initial conditions imposed by the StAW on the $Q_n$ variables but not on the  $\Pi_n$ variables. As a consequence, the non-null value  of $\langle\xi(t)\rangle$ is directly correlated to the initial conditions imposed on the $\Pi_n$ variables.
To recover that the average value of the stochastic force cancels, we impose that $\Omega_\nex t_0= 0[\pi]$: this corresponds to switching on  the StAW force at an extremum of the force.

\section{ The effective crystalline potential} 
\label{sec:potential}

The effective crystalline potential $\Phi_{\rm {eff}}(x)$ reads: 
\begin{eqnarray}
\Phi_{\rm {eff}}(x) &=&  \Phi_0(x) + \Delta \Phi_0(x),\\ 
\mbox{with  }  \Delta \Phi_0(x) &=& - \sum_n \frac{1}{ 2 \omega_n^2}   \Psi_n(x) \bar{\Psi}_n(x). \label{eq:dphi}
\end{eqnarray}

Using Eq.~\eqref{eq:Psinex0}, $\Delta \Phi_0(x)$ writes:
\begin{equation}
 \Delta \Phi_0(x)=- \sum_n \frac{1}{ 2 \omega_n^2}   \alpha_n(x) \bar{\alpha}_n(x).
\end{equation}
$\Delta \Phi_0(x)$ is then a periodic function of the lattice.  $\Delta \Phi_0(x)$ physically corresponds to the modification of the potential seen by the ad-atom induced by the auto-coherent  interaction between the substrate atoms and the ad-atom at position $x$.  Such interaction also appears in the memory kernel. Actually, both terms $\Delta \Phi_0(x)$ and the memory kernel derive from the integration by parts of the third term of Eq.~\eqref{eq:Qnt1} leading to Eq.~\eqref{eq:Qnt2}. The term $\Delta \Phi_0(x)$ derived from the third term of Eq.~\eqref{eq:Qnt2}, corresponds to the static and instantenous modification of the substrate variables due to the presence of the ad-atom at position $x$, while the memory kernel derived from the fourth term of Eq.~\eqref{eq:Qnt2}, corresponds to the retarded effects, i.e. how the past positions of the ad-atom influence the substrate positions at present.     
Both quantities $\Delta \Phi_0(x)$ and $\gamma(x,x',t-t')$ can be related introducing the function $\Theta(x,x',t-t')$:
\begin{eqnarray*} 
 \Theta(x,x',t-t')& =&   \sum_n \frac{ \cos(\omega_n (t-t')) }{ \omega_n^2} \bar{\Psi}_n(x')   \frac{d\Psi_n}{dx}(x), \nonumber\\
  \frac{d\Delta \Phi_0(x)}{dx}   &=&  - \frac{1}{2} \left[\Theta(x,x,0) + \bar{\Theta}(x,x,0)\right],  \\
\gamma(x,x',t-t') &=& \frac{ \partial \Theta } {\partial x'} (x,x',t-t'). \nonumber
\end{eqnarray*}
An explicit expression of the spatial dependence of $\Delta \Phi_0(x)$  can be obtained using the particular inter-atomic potential (Eq.~\eqref{eq:pot_phi}), and the $\Psi_n$ expression of Eq.~\eqref{eq:Psinex}, in Eq.~\eqref{eq:dphi}: 
\begin{widetext}
\begin{eqnarray} 
\Delta \Phi_0(x)&=& - \frac{ V_0^2 } { 2 \sigma^2 M}    \bigg(  \left[ \cosh^2(\frac{r}{\sigma}) +\cosh^2(\frac{r-a}{\sigma}) \right]   \sum_n \frac{1}{ \omega_n^2 
\left[ \cosh( \frac{a}{\sigma}) - \cos(k_n a) \right]^2 } \label{eq:pot_modif}  \\
&&- 2 \cosh(\frac{r}{\sigma}) \cosh(\frac{r-a}{\sigma})  \sum_n \frac{\cos(k_n a) }{ \omega_n^2\left[ \cosh( \frac{a}{\sigma}) - \cos(k_n a) \right]^2 } \bigg),  \nonumber 
\end{eqnarray}
\end{widetext}
where the two sums are only numerical factors independant of $x$. We recover in Eq.~\eqref{eq:pot_modif} that $\Delta \Phi_0(x)$ is a periodic function of the lattice.

\section{Conclusion} 

We have studied the diffusion of an ad-atom on a substrate submitted to a StAW. We found that the ad-atom motion is governed by a Generalized Langevin equation: 
\begin{eqnarray}
m \frac{d^2 x} {dt^2}\nonumber &=&  - \frac{d\Phi_{\rm{eff}}}{d x}(x)  + \xi(t)   
-   \int_{t_0}^t \gamma(x,x',t-t') \frac{dx}{dt}(t')  dt'  \nonumber \\
&&	  +   F_{SAW}(x,t)  \label{eq:endend}
\end{eqnarray}
We have characterized each of the terms involved in this equation and have given them their analytical expression and most of the time, an explicit expression. A key-result is the expression of the force $F_{SAW}$ induced by the StAW as a function of $x$ and $t$.  $F_{SAW}$ essentially varies as $\cos(k_\nex x) \cos(\Omega_\nex t)$ where $k_\nex$ and $\Omega_\nex$ are the spatial and angular frequencies of the StAW. However, a deeper analysis exhibits that this force also varies on the crystalline substrate lattice scale.
The next paper of this series is devoted to the study of the solutions of the equation Eq.~\eqref{eq:endend}.

\end{document}